Comment on "Doubly hybrid density functional xDH-PBE0 from a parameter-free global hybrid model PBE0" [J. Chem. Phys. 136, 174103 (2012)]

*Manoj K. Kesharwani,[1] Sebastian Kozuch,[2,*] and Jan M.L. Martin[1,*]*

(1) Department of Organic Chemistry, Weizmann Institute of Science, 76100 Reḥovot, Israel. Email: gershom@weizmann.ac.il. FAX: +972 8 9343029

(2) Department of Chemistry, Ben-Gurion University of the Negev, 84105 Beer Sheva, Israel. Email: kozuch@bgu.ac.il

Double hybrid functionals (see Ref.[1,2] for a review) have recently emerged as an interesting 'third way' option between DFT (density functional theory) and high-level *ab initio* methods: their accuracy approaches the latter at only moderate cost increase over the former, especially when RI (resolution of the identity[3,4]) is applied in the MP2 (2nd-order Møller-Plesset) phase.

There are two basic implementations of double hybrids in the literature. In the original Grimme approach[5] (denoted gDH throughout the paper), a Kohn-Sham calculation is carried out with a fraction $c_X$ of Hartree-Fock exchange and $(1-c_X)$ of DFA (density functional approximation) exchange, plus DFA correlation damped by a factor $c_{C,DFA}$. Then the MP2 correlation energy is evaluated in the basis of the Kohn-Sham orbitals obtained, scaled, and added to the energy total. In the more general DSD form, the energy is given by:

$$E = E_{NTVJ} + c_X E_{X,HF} + (1-c_X) E_{X,DFA} + c_{C,DFA} E_{C,DFA} + c_{2ss} E_{2ss} + c_{2ab} E_{2ab} + c_{disp} E_{dispersion} \quad (1.1)$$

where $E_{NTVJ}$ stands for the sum of nuclear repulsion, kinetic energy, electron-nuclear attraction, and Coulomb energies; $E_{X,HF}$ is the Hartree-Fock like exchange energy; $E_{X,DFA}$ the density functional exchange energy, $E_{2ss}$ is the same-spin MP2-like correlation energy; $E_{2ab}$ its opposite-spin counterpart; and $E_{dispersion}$ is an empirical dispersion model.[6] The orbitals are evaluated self-consistently for the given values of $c_X$ and $c_{C,DFA}$: typically, $c_X$ falls in the 50–70% range, and $c_{C,DFA}$ is considerably less than unity.

In the XYG3[7] or xDH[8] approach, on the other hand, the orbitals used for the evaluation of all terms in Eq. (1) are evaluated for a standard hybrid with full DFA correlation (i.e., $c_{C,DFA}$=100%), and with $c_{X,HF}$ as appropriate for a conventional hybrid DFA, i.e. typically in the 20-25% range.

It has been argued that the xDH approach is more appropriate, based on the fact that the orbitals employed in eq. (1) would be more realistic Kohn-Sham orbitals for the system.[7] On the other hand, the very low RMSD values over extensive training sets obtained for functionals like B2GP-PLYP[9] and DSD-PBEP86[10] speak for themselves. Goerigk and Grimme, in Section 5.3 of their GMTKN30 benchmark paper,[11] address the issue specifically for XYG3 vs. B2PLYP[5] and B2GP-PLYP. They consider the average occupied-virtual gap in the orbital energies and find it to be relatively indifferent to the percentage of GGA correlation, but quite sensitive to the percentage of HF exchange, with smaller percentages yielding smaller band gaps (i.e., perturbation denominators), and hence higher effective percentages of MP2-like correlation.[12]

The purpose of the present work is to make a head-to-head comparison, for the same training set, of the performance of DSD functionals and what we will term xDSD functionals, which are the same forms as in Eq.(1) but with orbitals for all terms obtained as in xDH functionals, i.e., from a hybrid GGA calculation with a set percentage of HF exchange and undamped correlation.

Details and references for the (m)GGA exchange functionals attempted here are given in the Supporting Information.[13] Both the D2[14] and D3BJ[15,16] dispersion models were considered.

All calculations were carried out using the Gaussian 09 Rev. D.01 program system[17] running on the Faculty of Chemistry computing farm at the Weizmann Institute of Science.

Six reference datasets were used, which constitute an updated version of the training set used in Refs.[10,18] They cover atomization energies, main group barrier heights, noncovalent interactions, and late transition metal catalysis. Further details and references are given in the Supporting Information. The arithmetic average of all six RMSDs, AveRMSD, is used as the principal metric. Key results and optimized functional parameters can be found in Table S1 in the Supporting Information, detailed results in an Excel workbook there.

For xDH-PBE0, AveRMSD is 3.04 kcal/mol; in contrast, the DSD-PBEPBE-D3BJ functional has just 1.512 kcal/mol (compared to 1.62 in Refs.[10,18]). Comparison of the parameters reveals that, while the DFT and opposite-spin MP2 coefficients are not too different, xDH-PBE0 has a much larger fraction of HF exchange in the final result (83.4%, vs. 68% for DSD-PBE-D3BJ). Of course, $c_{disp}=c_{2ss}=0$ in xDH-PBE0; relaxing these constraints and reoptimizing, AveRMSD drops to 1.60 kcal/mol, not significantly different from DSD-PBEPBE-D3BJ. The optimal $c_{2ss}$ is quite small: constraining it to zero increases AveRMSD by just 0.06 kcal/mol.

The underlying orbitals in xDH-PBE0 and xDSD-PBE0-D3BJ are those of the PBE0 functional[19] with $c_X=1/4$. We then proceeded to consider additional data points (xDSD-PBE$_x$-D2 in Table S1) where the orbitals were obtained with $c_X = \{0, 1/3, 3/8, 1/2, 0.68, 3/4, 1\}$. Predictably, for $c_X = 0$ or 1 the AveRMSD is elevated (albeit still lower than xDH-PBE0); less obvious perhaps is that the lowest AveRMSD values are obtained for $c_X=1/2$, with $c_X=0.68$ marginally higher.

In fact, xDSD-PBE$_{68}$-D3BJ performs somewhat *better* (AveRMSD=1.37 kcal/mol) than DSD-PBEPBE-D3BJ (AveRMSD=1.51 kcal/mol), which has the same $c_x$ but damped DFT correlation in the orbitals. It is worth noting that the $c_x$ parameters of the two functionals are basically identical: $c_X$ = 0.683 for xDSD vs. 0.680 for DSD. This suggests that the relatively poor performance of xDH-PBE0 is not due to the choice of reference orbitals but to the absence of both same-spin MP2 and a dispersion correction (which are known[20,21] to contain very similar information).

What about other functionals? Let us consider orbitals with $c_X$=1/2 for the xDSD forms, and the simpler D2 dispersion correction, i.e., a simple multivariate-linear optimization. xDSD-S$_{50}$VWN5-D2, where the underlying DFT functional is just a local density approximation, puts in a surprisingly good performance (AveRMSD=1.39 kcal/mol). xDSD-B$_{50}$B95-D2, xDSD-B$_{50}$HLYP, xDSD-TPSS$_{50}$ all perform somewhat worse. The winner is xDSD-PBE$_{50}$P86-D2 with just AveRMSD=1.34 kcal/mol. Substituting the improved D3BJ dispersion correction and reoptimizing to obtain xDSD-PBE$_{69}$P86-D3BJ, AveRMSD can be lowered further to 1.22 kcal/mol, compared to 1.36 kcal/mol for DSD-PBEP86-D3BJ. Again, the parameters are now fairly similar between DSD and xDSD variants.

xDSD appears to have a slight edge over DSD for the (x)DSD-PBE and (x)DSD-PBEP86 combos, less so for (x)DSD-PBEhP95. Some caution against over-analysis is due here as the differences are arguably comparable to the remaining uncertainties in the reference data. Eliminating same-spin correlation typically leads to a small increase in AveRMSD and an increase in the prefactor for the dispersion correction, consistent with the repeatedly noted[20,21] similarity between dispersion and same-spin MP2 correlation energy. Eliminating both leads to a significant deterioration in AveRMSD; one or the other needs to be left in.

Finally, we considered how transferable the orbitals are. For instance, if we plug in PBE0 orbitals into xDSD-PBE$_{25}$P86 rather than using PBE$_{25}$P86 converged orbitals, this affects AveRMSD by a paltry 0.01 kcal/mol. This indicates that the results are fairly insensitive to the particular exchange and correlation forms employed for those "reference" orbitals.

Summing up, we have compared the performance of Grimme type gDH/DSD and Zhang-Xu-Goddard type xDH/xDSD forms for double hybrids. In the gDH and DSD forms, KS orbitals with elevated HF exchange and damped DFT correlation are used, while in the xDH and xDSD forms, the KS orbitals are obtained from a conventional hybrid functional with undamped DFT correlation. Generally, the difference in performance between gDSD and xDSD functionals is very small, slightly favoring xDSD. Augmentation of the xDH form with either same-spin MP2 correlation or a dispersion correction markedly improves performance. Best xDSD results appear to be obtained for orbitals obtained with "exact exchange" fractions in the 50-70% range. The orbitals for xDSD appear to be fairly transferable between different correlation functionals.

ACKNOWLEDGMENT. This research was supported by the Lise Meitner-Minerva Center for Computational Quantum Chemistry, by the Weizmann Yeda-Sela Initiative, and by the Helen and Martin Kimmel Center for Molecular Design. The authors would like to thank Dr. Douglas J. Fox of Gaussian, Inc. for helpful suggestions concerning the l608 module.

REFERENCES

[1] L. Goerigk and S. Grimme, Wiley Interdiscip. Rev. Comput. Mol. Sci. **4**, 576 (2014).

[2] I.Y. Zhang and X. Xu, Int. Rev. Phys. Chem. **30**, 115 (2011).

[3] F. Weigend and M. Häser, Theor. Chem. Acc. **97**, 331 (1997).

[4] R.A. Kendall and H.A. Früchtl, Theor. Chem. Acc. **97**, 158 (1997).

[5] S. Grimme, J. Chem. Phys. **124**, 034108 (2006).

[6] S. Grimme, Wiley Interdiscip. Rev. Comput. Mol. Sci. **1**, 211 (2011).

[7] Y. Zhang, X. Xu, and W.A. Goddard, Proc. Natl. Acad. Sci. **106**, 4963 (2009).

[8] I.Y. Zhang, N.Q. Su, E. a. G. Brémond, C. Adamo, and X. Xu, J. Chem. Phys. **136**, 174103 (2012).

[9] A. Karton, A. Tarnopolsky, J.-F. Lamère, G.C. Schatz, and J.M.L. Martin, J. Phys. Chem. A **112**, 12868 (2008).

[10] S. Kozuch and J.M.L. Martin, Phys. Chem. Chem. Phys. **13**, 20104 (2011).

[11] L. Goerigk and S. Grimme, J. Chem. Theory Comput. **7**, 291 (2011).

[12] M. Ernzerhof and J.P. Perdew, J. Chem. Phys. **109**, 3313 (1998).

[13] See EPAPS Document Number EPAPS-JCPSA6-Xxx-Yyyyyy for details of exchange and correlation functionals, details of training sets, Table S1, and an Excel spreadsheet containing detailed results for all the training sets and functionals. For More Information on EPAPS, See Http://www.aip.org/pubservs/epaps.html .

[14] S. Grimme, J. Comput. Chem. **27**, 1787 (2006).

[15] S. Grimme, J. Antony, S. Ehrlich, and H. Krieg, J. Chem. Phys. **132**, 154104 (2010).

[16] S. Grimme, S. Ehrlich, and L. Goerigk, J. Comput. Chem. **32**, 1456 (2011).

[17] M.J. Frisch, G.W. Trucks, H.B. Schlegel, G.E. Scuseria, M.A. Robb, J.R. Cheeseman, G. Scalmani, V. Barone, B. Mennucci, G.A. Petersson, H. Nakatsuji, M. Caricato, X. Li, H.P. Hratchian, A.F. Izmaylov, J. Bloino, G. Zheng, J.L. Sonnenberg, M. Hada, M. Ehara, K. Toyota, R. Fukuda, J. Hasegawa, M. Ishida, T. Nakajima, Y. Honda, O. Kitao, H. Nakai, T. Vreven, J. Montgomery, J. A., J.E. Peralta, F. Ogliaro, M. Bearpark, J.J. Heyd, E. Brothers, K.N. Kudin, V.N. Staroverov, R. Kobayashi, J. Normand, K. Raghavachari, A.P. Rendell, J.C. Burant, S.S. Iyengar, J. Tomasi, M. Cossi, N. Rega, M.J. Millam, M. Klene, J.E. Knox, J.B. Cross, V. Bakken, C. Adamo, J. Jaramillo, R. Gomperts, R.E. Stratmann, O. Yazyev, A.J. Austin, R. Cammi, C. Pomelli, J.W. Ochterski, R.L. Martin, K. Morokuma, V.G. Zakrzewski, G.A. Voth, P. Salvador, J.J. Dannenberg, S. Dapprich, A.D. Daniels, Ö. Farkas, J.B. Foresman, J. V. Ortiz, J. Cioslowski, and D.J. Fox, Gaussian 09 Rev.D01 (Gaussian, Inc., Wallingford, CT, 2014).

[18] S. Kozuch and J.M.L. Martin, J. Comput. Chem. **34**, 2327 (2013).

[19] C. Adamo and V. Barone, J. Chem. Phys. **110**, 6158 (1999).

[20] J.M.L. Martin, J. Phys. Chem. A **117**, 3118 (2013).

[21] S. Kozuch, D. Gruzman, and J.M.L. Martin, J. Phys. Chem. C **114**, 20801 (2010).

# Comment on "Doubly hybrid density functional xDH-PBE0 from a parameter-free global hybrid model PBE0" [J. Chem. Phys. 136, 174103 (2012)]

*Manoj K. Kesharwani,[1] Sebastian Kozuch,[2,*] and Jan M.L. Martin[1,*]*

(1) Department of Organic Chemistry, Weizmann Institute of Science, 76100 Reḥovot, Israel. Email: gershom@weizmann.ac.il. FAX: +972 8 9344142

(2) Department of Chemistry, Ben-Gurion University of the Negev, 84105 Beer Sheva, Israel. Email: kozuch@bgu.ac.il

**Supporting Information**

**DFT exchange and correlation components**

The exchange functionals considered here are LDA (Slater), the GGAs PBE,[1] PBEh,[2] and B88,[3] and the meta-GGA TPSS,[4] while the correlation functionals considered are the VWN5 parametrization[5] of LDA, P86,[6] PBE,[1] LYP,[7] B95,[8] and TPSS.

**Reference datasets**

A slightly updated version of the training set from Refs.[9,10] has been used. In these older studies, we employed the W4-08 dataset for atomization energies;[11] the revised DBH24 barrier heights;[12,11] the "mindless benchmark" (artificial structures) of Korth and Grimme;[13]

the revised S22 noncovalent interactions dataset[14,15] prototype insertion reactions at bare Pd of Quintal et al;[16] and the Zhao-Truhlar model model[17] for the Grubbs catalyst. In the present work, W4-08 was replaced with the larger and more recent W4-11 dataset[18] and the energetics for the bare-Pd and Grubbs benchmarks were revised from a recent explicitly correlated *ab initio* study.[19]

[1] J. Perdew, K. Burke, and M. Ernzerhof, Phys. Rev. Lett. **77**, 3865 (1996).

[2] M. Ernzerhof and J.P. Perdew, J. Chem. Phys. **109**, 3313 (1998).

[3] A.D. Becke, Phys. Rev. A **38**, 3098 (1988).

[4] J. Tao, J. Perdew, V. Staroverov, and G. Scuseria, Phys. Rev. Lett. **91**, 146401 (2003).

[5] S.H. Vosko, L. Wilk, and M. Nusair, Can. J. Phys. **58**, 1200 (1980).

[6] J.P. Perdew, Phys. Rev. B **33**, 8822 (1986).

[7] C. Lee, W. Yang, and R.G. Parr, Phys. Rev. B **37**, 785 (1988).

[8] A.D. Becke, J. Chem. Phys. **104**, 1040 (1996).

[9] S. Kozuch and J.M.L. Martin, Phys. Chem. Chem. Phys. **13**, 20104 (2011).

[10] S. Kozuch and J.M.L. Martin, J. Comput. Chem. **34**, 2327 (2013).

[11] A. Karton, A. Tarnopolsky, J.-F. Lamère, G.C. Schatz, and J.M.L. Martin, J. Phys. Chem. A **112**, 12868 (2008).

[12] J. Zheng, Y. Zhao, and D.G. Truhlar, J. Chem. Theory Comput. **5**, 808 (2009).

[13] M. Korth and S. Grimme, J. Chem. Theory Comput. **5**, 993 (2009).

[14] P. Jurecka, J. Sponer, J. Cerný, and P. Hobza, Phys. Chem. Chem. Phys. **8**, 1985 (2006).

[15] T. Takatani, E.G. Hohenstein, M. Malagoli, M.S. Marshall, and C.D. Sherrill, J. Chem. Phys. **132**, 144104 (2010).

[16] M.M. Quintal, A. Karton, M.A. Iron, A.D. Boese, and J.M.L. Martin, J. Phys. Chem. A **110**, 709 (2006).

[17] Y. Zhao and D.G. Truhlar, J. Chem. Theory Comput. **5**, 324 (2009).

[18] A. Karton, S. Daon, and J.M.L. Martin, Chem. Phys. Lett. **510**, 165 (2011).

[19] M.K. Kesharwani and J.M.L. Martin, Theor. Chem. Acc. **133**, 1452 (2014).

Table S1: Selected DH/DFT coefficients (see eq. 1.1) and their average RMSD for the training sets. The "x" functionals use a conventional hybrid functional to obtain the KS orbitals, with the value in subscript after the exchange DFT indicating the amount of exact exchange in the KS orbitals.

| Functional | $c_X$ | $c_{C,DFA}$ | $c_{2ab}$ | $s_6$ | $c_{2SS}$ | $a_2$ | AveRMSD |
|---|---|---|---|---|---|---|---|
| xDH-PBE0 | 0.834 | 0.529 | 0.543 | 0.000 | 0.000 | N/A | 3.037 |
| DSD-PBEPBE-D3BJ | 0.680 | 0.490 | 0.550 | 0.780 | 0.130 | 6.1 | 1.512 |
| xDSD-PBE0-D3BJ | 0.689 | 0.616 | 0.384 | 0.746 | 0.071 | 6.2 | 1.603 |
| xDOD-PBE0-D3BJ | 0.654 | 0.642 | 0.380 | 0.923 | 0.000 | 6.2 | 1.666 |
| xDSD-PBE$_{68}$-D3BJ | 0.683 | 0.475 | 0.564 | 0.735 | 0.141 | 6.2 | 1.367 |
| xDSD-PBE$_{pureGGA}$-D2 | 0.557 | 0.783 | 0.218 | 0.509 | 0.010 | N/A | 2.353 |
| xDSD-PBE0-D2 | 0.690 | 0.612 | 0.390 | 0.384 | 0.071 | N/A | 1.644 |
| xDSD-PBE$_{33.3}$-D2 | 0.704 | 0.573 | 0.432 | 0.362 | 0.090 | N/A | 1.563 |
| xDSD-PBE$_{37.5}$-D2 | 0.711 | 0.553 | 0.456 | 0.352 | 0.098 | N/A | 1.507 |
| xDSD-PBE$_{50}$-D2 | 0.713 | 0.509 | 0.512 | 0.339 | 0.122 | N/A | 1.438 |
| xDSD-PBEP$_{68}$-D2 | 0.693 | 0.461 | 0.578 | 0.342 | 0.157 | N/A | 1.472 |
| xDSD-PBE$_{75}$-D2 | 0.681 | 0.446 | 0.601 | 0.346 | 0.173 | N/A | 1.563 |
| xDSD-PBE$_{100}$-D2 | 0.608 | 0.432 | 0.651 | 0.384 | 0.202 | N/A | 2.429 |
| DSD-PBEP86-D3BJ | 0.690 | 0.440 | 0.520 | 0.460 | 0.230 | 5.6 | 1.360 |
| xDSD-PBE$_{69}$P86-D3BJ | 0.697 | 0.416 | 0.553 | 0.446 | 0.211 | 5.6 | 1.224 |
| xDSD-PBE$_{50}$P86-D2 | 0.731 | 0.451 | 0.489 | 0.226 | 0.210 | N/A | 1.340 |
| xDSD-PBE$_{69}$P86-D2 | 0.701 | 0.414 | 0.553 | 0.249 | 0.239 | N/A | 1.327 |
| DSD-PBEhB95-D3BJ | 0.660 | 0.550 | 0.470 | 0.580 | 0.090 | 6.2 | 1.473 |
| xDSD-PBE$_{66}$B95-D3BJ | 0.667 | 0.543 | 0.476 | 0.483 | 0.099 | 6.2 | 1.435 |
| xDSD-PBE$_{50}$B95-D2 | 0.671 | 0.582 | 0.425 | 0.243 | 0.059 | N/A | 1.412 |
| xDSD-PBE$_{66}$B95-D2 | 0.664 | 0.537 | 0.485 | 0.258 | 0.091 | N/A | 1.458 |
| xDSD-S$_{50}$VWN5-D2 | 0.746 | 0.380 | 0.501 | 0.353 | 0.166 | N/A | 1.391 |
| xDSD-B$_{50}$B95-D2 | 0.676 | 0.600 | 0.424 | 0.345 | 0.082 | N/A | 1.616 |
| xDSD-B$_{50}$LYP-D2 | 0.747 | 0.549 | 0.441 | 0.371 | 0.354 | N/A | 1.836 |
| xDSD-TPSS$_{50}$-D2 | 0.779 | 0.446 | 0.552 | 0.217 | 0.271 | N/A | 1.731 |
| xDSD-PBE$_{25}$-P86-D2 | 0.709 | 0.543 | 0.368 | 0.256 | 0.154 | N/A | 1.599 |
| PBE0 KS orbitals into xDSD-PBEP86 | 0.708 | 0.544 | 0.367 | 0.256 | 0.155 | N/A | 1.610 |